\documentstyle[12pt]{article}
\begin{document}
\begin{center}
{\large  Generalization of Weierstrassian Elliptic Functions to }
${\bf R}^{n}$ \\[20mm]
{\large Cihan Sa\c{c}l{\i}o\u{g}lu}\footnote{Permanent address:
Bo\u{g}azi\c{c}i University, Department of Physics, 80626 Istanbul,
Turkey.}\\[2mm]
 T\"{U}B\.{I}TAK - Marmara Research Center \\
 Research Institute for Basic Sciences \\
 Department of Physics \\
 41470 Gebze, Turkey \\[5mm]
\today
\begin{abstract}
The Weierstrassian $\wp, \zeta$  and $\sigma $ functions are generalized
to ${\bf R}^{n}$. The $n=3$ and $n=4$ cases have already been used in
instanton solutions which may be interpreted as explicit realizations
of spacetime foam and the monopole condensate, respectively.
The new functions satisfy higher dimensional versions
of the  periodicity properties and Legendre's relations obeyed by their
familiar complex counterparts.  For $n=4$, the construction reproduces
functions found earlier by Fueter using quaternionic methods.
Integrating over lattice points along all directions but two,
one recovers the original Weierstrassian elliptic functions.
\end{abstract}
\end{center}
\vspace{2cm}

\pagebreak

Kaluza-Klein, Supergravity, String and Superstring
theories and Super p-branes all involve dimensions beyond four,
which then have to be compactified.  The compactification results in
lattices in the higher dimensions \cite{l}. It might therefore be of
interest to consider generalizations of doubly periodic functions
to arbitrary $ {\bf R}^{n} $. The $ {\bf R}^{4} $ case has already been
treated by Fueter \cite{f}, who succeeded in obtaining quaternionic
analogues of Weierstrassian elliptic functions. In this note, we will
present a unique and straightforward extension of Fueter's results to
$ {\bf R}^{n} $. This general construction yields Fueter's functions
in $ n=4 $ without requiring the use of quaternions. Furthermore, when
one takes infinitesimal lattice spacings in all but two
lattice directions and integrates over these points one recovers the
usual Weierstrassian functions.

Apart from possible future physics applications of n-tuply
periodic functions, it is worth noting that (quasi)periodic classical
solutions of Gravity and Yang-Mills theory have already been
considered. Thus, inspired by Rossi's observation \cite{r} that a singly-
periodic instanton configuration represents a BPS monopole \cite{B}, Gursey
and Tze \cite{G} used Fueter's hyperelliptic functions to construct
a solution with one Yang-Mills instanton per spacetime cell.  This
solution may be related to the monopole condensate \cite{m} or to the
Copenhagen vacuum \cite{n},  which is also based on a periodic arrangement of
the theory's solitons.  Hawking, on the other hand, argued \cite{H}
that metrics with a single gravitational instanton per unit cell
make the dominant contribution to the path integral of Einstein's
gravity.  An adaptation of Weierstrassian $\sigma$ and $ \zeta $
functions to three dimensions \cite{N-S} along the lines presented here
leads to just such a self-dual solution, providing an explicit example of
a "spacetime foam" based on Gibbons-Hawking multicenter metrics \cite{g}.

We start by examining how the infinite sums defining the Weierstrassian
functions are made to converge. These are

\begin{equation}
\wp(z) = \frac{1}{z^2} +   \sum_{\omega\neq 0} \{ \frac{1}{(z-\omega)^2}-
\frac{1}{\omega^2}\},
\end{equation}

\begin{equation}
\zeta(z) =-\int\wp(z)dz = \frac{1}{z} +
\sum_{\omega\neq 0} \{ \frac{1}{(z-\omega)}+
\frac{1}{\omega}+\frac{z}{\omega^2} \},
\end{equation}
\begin{equation}
\rho(z)\equiv\ln\sigma(z)=\int dz\,\zeta(z)=\ln(z)+
\sum_{\omega\neq 0}\{\ln(z-\omega)-
\ln(-\omega)+\frac{z}{\omega}+\frac{z^2}{2\omega^2}\},
\end{equation}
or,
\begin{equation}
\rho(z)= \ln(z) +\sum_{\omega\neq 0}\{\ln(1-\frac{z}{\omega})+
\frac{z}{\omega}
+\frac{z^2}{2\omega^2}\}
\end{equation}
In the above $\omega=n_{1}\omega_{1}+n_{2} \,\omega_{2}$, with
$(\omega_{1},\omega_{2})$ defining the basic lattice cell $\cal{C}$.
The parentheses $\{\}$ ensure that each term in the sum is absolutely
convergent; the series become meaningless if the parentheses are
broken up and the terms they contain are separately summed.  An
efficient method for studying the convergence properties of (1)-(3)
is provided by an integral test in which the double sum is replaced
by $\int_{|\omega|_{min}}^{\infty}d|\omega||\omega|$. This reveals why
one, two and three subtraction terms are needed in (1), (2) and (3),
respectively.  Employing a simple dimensional argument, one can proceed
further and determine the precise forms of the subtraction terms,
given only the first term in the sum (3), for example. To do this,
we note (a)  $\rho$ is dimensionless (assigning the dimension of length to
the coordinate z, say); (b) in the power series expansion of (4), the
highest power of $\omega$ permitted by convergence requirements is
$\omega^{-3}$.  One then realizes that the subtraction terms following
$\ln(1-\frac{z}{\omega})$ must be chosen so that $\rho(z)$ has the
expansion
\begin{equation}
\rho(z)=\ln z+\sum_{\omega\neq 0} O(\frac{z^3}{\omega^3})+...
\end{equation}
Hence using the power series
\begin{equation}
\ln(1-z/\omega)=-z/\omega-z^2/2\omega^2 -z^3/3\omega^3-....
\end{equation}
one chooses the three subtraction terms
\begin{equation}
-\ln(-\omega)+z/\omega+z^2/2\omega^2
\end{equation}
to arrive at the form (5).  This will be the key in constructing
${\bf R}^n$
analogues of $\rho(z).$

We will later need the well known facts that $\wp(z)$ is doubly
periodic while  $\zeta(z)$ and $\sigma(z)$ are quasiperiodic with the
transformation properties
\begin{equation}
\zeta(z+\omega_{1,2})=\zeta(z)+\eta_{1,2}
\end{equation}
and
\begin{equation}
\sigma(z+\omega_{1,2}) =-\sigma(z)\,exp\,\eta_{1,2}\,(z+\omega_{1,2}/2)
\end{equation}
where $\eta_{1,2}=2\zeta(\omega_{1,2}/2)$.

We also record the so-called Legendre's relations
\begin{equation}
\oint_{\partial\cal{C}}\zeta(z)\,dz=2\pi i
\end{equation}
and
\begin{equation}
\oint_{\partial\cal{C}}\zeta(z)\,dz=2\pi i=
\eta_{1}\omega_{2}-\eta_{2} \,\omega_{1}.
\end{equation}

We first rewrite (10) in the rather unconventional form reminiscent
of Gauss's theorem
\begin{equation}
\int\!\!\int_{\cal{C}}dV\,\nabla^{2}\rho(z)=\oint_{\partial\cal{C}}
d\vec{\sigma}\cdot\vec{\nabla}\rho(z)=2\pi,
\end{equation}
where $dV=dxdy$ and the "surface element" $d\vec{\sigma}=\hat{n}dl$.
Here $dl$ is the arclength and $\hat{n}$ a unit normal vector
pointing outwards on $\partial\cal{C}$.

Let us next introduce the lattice basis vectors $q^{(a)}_{\mu}$, where
$a,\mu=1,...,n$.  The volume of the unit cell $ \cal{C}$ is given by
\begin{equation}
V_{n}=\frac{1}{n!}\epsilon_{a_{1}...a_{n}}
\epsilon^{\mu_{1}...\mu_{n}}q^{(a_{1})}_{\mu_{1}}...q^{(a_{n})}_{\mu_{n}}.
\end{equation}
The basis vectors of the reciprocal lattice obey
\begin{equation}
r^{(a)}_{\mu}q^{(b)}_{\mu}\equiv
r^{(a)}\cdot q^{(b)}=\delta^{ab}.
\end{equation}
They are obtained from the $q^{(a)}_{\mu}$ via
\begin{equation}
r^{(a_{1})}_{\mu_{1}}=\frac{1}{(n-1)!V_{n}}
\epsilon_{a_{1}...a_{n}}\epsilon^{\mu_{1}...\mu_{n}}
q^{(a_{2})}_{\mu_{2}}...q^{(a_{n})}_{\mu_{n}}.
\end{equation}
We now seek higher dimensional versions of (12) in the form
\begin{equation}
\int_{\cal{C}}dV_{n}\,\partial_{\mu}\partial_{\mu}\,\rho_{n}(x)=
\oint_{\partial\cal{C}}d\sigma_{\mu}\partial_{\mu}\,\rho_{n}(x)=
-\int d\omega_{n},
\end{equation}
where of course
\begin{equation}
\int d\omega_{n}\equiv \Omega_{n}=2\pi^{n/2}/\Gamma(n/2).
\end{equation}
In order to obtain the result (16), the function  $\rho_{n}(x)$, which is
 to serve as the analog of $\rho(z)$, should have the form
\begin{equation}
\rho_{n}(x)\propto G_{n}(x)+\sum ..\sum_{q\neq 0}
{G_{n}(x-q)+({\bf R}^{n}  \,harmonics)}.
\end{equation}
In (18),  $G_{n}(x)$ is the Green's function for ${\bf R}^{n}$ obeying
\begin{equation}
\partial_{\mu}\partial_{\mu}\,G_{n}(x)=-\Omega_{n}\delta(x_{1})...
\delta(x_{n}) \equiv -\Omega_{n}\delta(x)
\end{equation}
and
\begin{equation}
q=n_{1}q^{(1)}+...+n_{n}q^{(n)}.
\end{equation}
We have, of course
\begin{equation}
G_{n}(x-q)=\frac{1}{(x^{2}-2x\cdot q+q^{2})^{\frac{n-2}{2}}}=
\frac{1}{|x-q|^{n-2}} \    \; (n>2).
\end{equation}
The harmonics in (18) should now be chosen according to the general strategy
outlined between equations (4) and (8).  Thus in order to render convergent
the sum (18), whose integral counterpart contains terms behaving like
 $\int_{|q|_{min}}^{\infty}d|q||q|^{n-1}/|q|^{n-2}$, one again needs
3 subtraction terms. These are in fact nothing but the first three
terms in the MacLauren expansion of (21) for $|x|<<|q|_{min}$.  This
immediately leads to
\begin{equation}
\rho_{n}(x)=\frac{1}{(x^{2})^{\frac{n-2}{2}}}+\sum..\sum_{q\neq 0}
\{\frac{1}{|x-q|^{n-2}}-\frac{1}{|q|^{n-2}}[1+\frac{(n-2)}{q^{2}}
(q\cdot x+\frac{1}{2q^{2}}(n(q\cdot x)^2-q^{2}x^{2}))]\}
\end{equation}
Again, absolute convergence is attained only by considering each term
defined by the outermost parentheses as an indivisible unit.
Hence the terms proportional to $q\cdot x$ and to $1/2q^{2}$ cannot
be summed separately (in which case they would appear to give zero
by symmetry!) anymore than the $\omega^{-1}$ term in (2) can.  Note
that the subtraction terms are indeed harmonics as anticipated in (18).

This is perhaps an appropriate point for comparing our results with
Fueter's.  Fueter introduces the unit quaternions ${\bf e}_{\mu}$ and
$ \overline{{\bf e}}_{\mu}$ corresponding to $ (I, {\bf e}_{i})$ and
$(I,-{\bf e}_{i})$, respectively.  We have,  as usual,
${\bf e}_{i} {\bf e}_{j}
=-\delta_{ij}+\epsilon_{ijk}{\bf e}_{k}$, the indices $i,j, k$ running
from 1 to 3. Using these, one defines {${\bf x}=x_{\mu}{\bf e}_{\mu},\,
\overline{{\bf x}}=x_{\mu}\overline{{\bf e}}_{\mu}, D={\bf e}_{\mu}
\partial_{\mu}$} and $ \overline{D}=\overline{{\bf e}}_{\mu}\partial_{\mu}$.
One then has $x^{2}=\overline{{\bf x}}{\bf x} $ and $\overline{D}D=
\partial_{\mu}\partial_{\mu}.$
Then, starting with the function
\begin{equation}
{\bf Z({\bf x})}=\frac{1}{{\bf x}}+\sum_{q\neq 0}\{\frac{1}{{\bf x-q}}+
\frac{1}{{\bf q}}+\frac{1}{{\bf q}}{\bf x}\frac{1}{{\bf q}}+
\frac{1}{{\bf q}}{\bf x}\frac{1}{{\bf q}}{\bf x}\frac{1}{{\bf q}}+
\frac{1}{{\bf q}}{\bf x}\frac{1}{{\bf q}}{\bf x}
\frac{1}{{\bf q}}{\bf x}\frac{1}{{\bf q}}\},
\end{equation}
where ${\bf q}=n_{a}{\bf q}^{(a)}=n_{a}q^{(a)}_{\mu}{\bf e}_{\mu},$ the
$q^{(a)}$ being lattice vectors, one arrives at the quaternionic version
of Weierstrass's $\zeta (z)$ via
\begin{equation}
{\bf \zeta^{F}({\bf x})}=\partial_{\mu}\partial_{\mu}{\bf Z({\bf x})}.
\end{equation}
We will see later that  (24) indeed transforms the same way as $\zeta (z)$
under lattice shifts. The function corresponding to $\ln \sigma (z)$ is
\begin{equation}
\rho({\bf x})= D{\bf Z({\bf x})}
\end{equation}
and, remarkably, it turns out to have the quaternion-free form
\begin{equation}
\rho (x)=\frac{1}{x^{2}}+\sum_{q\neq 0}\{\frac{1}{(x-q)^{2}}-
\frac{1}{q^{2}}-
\frac{2x\cdot q}{q^{4}}-\frac{1}{q^{6}}(4(q\cdot x)^{2}-q^{2}x^{2})\},
\end{equation}
coinciding with the n=4 case of (22).

Returning next to (12), we see that the analog of $\zeta (z)$ is the
n-gradient
\begin{equation}
\zeta^{(n)}_{\mu}(x)=\partial_{\mu}\,\rho_{n}(x).
\end{equation}
Taking n=4 and contracting both sides with the quaternion units
$\overline{{\bf e}}_{\mu},\,$ we recover the $ {\bf \zeta^{F}}$ function
of Fueter [2]
\begin{equation}
{\bf \zeta^{F}} =\overline{{\bf e}}_{\mu}\,\partial_{\mu}\,\rho_{4}
=\overline{D}\rho_{4}.
\end{equation}

Leaving the quaternionic special case aside, let us now
find the higher dimensional  analogs of (8) and (9). Since
\begin{equation}
\partial_{\mu}\partial_{\mu}\,\rho_{n}=\partial_{\mu}\,\zeta^{(n)}_{\mu}=-
\Omega_{n}\sum..\sum\delta (x-q)
\end{equation}
is a perfectly n-tuply periodic distribution, a shift by a lattice basis
vector can only change $\zeta^{(n)}_{\mu}$ by a constant vector.  Thus
\begin{equation}
\zeta^{(n)}_{\mu}(x+q^{(a)})=\zeta^{(n)}_{\mu}(x)+\eta^{(n)(a)}_{\mu}.
\end{equation}
This is indeed the transformation law for ${\bf \zeta^{F}}$ when n=4.
Putting $x=-q^{(a)}/2$ and noting that $\zeta^{(n)}_{\mu}(-x)=
-\zeta^{(n)}_{\mu}(x)$, we find
\begin{equation}
2\zeta^{(n)}_{\mu}(q^{(a)}/2)=\eta^{(n)(a)}_{\mu}.
\end{equation}
Integrating (30) and using $\rho_{n}(-x)=\rho_{n}(x)$, we obtain
\begin{equation}
\rho_{n}(x+q^{(a)})=\rho_{n}(x)+\eta^{(n)(a)}\cdot (x+q^{(a)}/2),
\end{equation}
where there is no sum on the index $(a)$ on the right hand side.
Using (30), we can also obtain the higher dimensional form of Legendre's
second relation (11).  Let us evaluate $\oint_{\partial\cal{C}}
d\sigma_{\mu}\zeta^{(n)}_{\mu}(x) $, where $ \partial\cal{C}$ is the
surface of the fundamental hyperparallelepiped.  Noting that the
normal to the hyperplane defined by  $\{q^{(a_{2})},..,q^{(a_{n})}\}$
is along the reciprocal vector $r^{(a_{1})}$ given in (15), we find
\begin{equation}
V_{n}\sum_{a=1}^{n}\eta^{(n)(a)}\cdot r^{(a)}=\Omega_{n}=
\frac{2\pi^{\frac{n}{2}}}{\Gamma(\frac{n}{2})}
\end{equation}
as the n-dimensional generalization of Legendre's second relation.

There is an interesting relationship between $\rho_{n}$ and $\rho_{n-1}$
which can be iterated until one finally arrives at the Weierstrassian
elliptic functions. It is simplest to describe this relationship in
the case of a rectangular lattice although it is generally valid.
Starting with $\rho_{n},$  one first chooses one lattice direction to
coincide with, say, the $x_{n}$ axis and
then lets the lattice spacing (in the same direction only) become
infinitesimal. Then,  integrating over $q_{n}$,
one obtains $\rho_{n-1}$ up to a multiplicative constant.
Repeating this procedure until $\rho_{3}$ is found, one finally has
\begin{equation}
-\int_{-\infty}^{\infty} dq_{3} \,\rho_{3}=\rho (z)+\rho (\overline{z})=
\int dz\,\zeta (z)+\int d\overline{z}\,\zeta (\overline{z}),
\end{equation}
where the remaining lattice basis vectors
$(\vec{q}^{\,(1)},\vec{q}^{\,(2)})$
provide the periods $(\omega_{1}, \omega_{2})$.  Thus the $\rho_{n}$
presented here are indeed intimately connected with Weierstrassian
functions.  We also observe that the split into an analytic and
anti-analytic part is a feature of two dimensions, not shared for
example by Fueter's quaternionic functions.

Finally, the reader may wonder how the ${\bf R^{n}}$ counterpart of
$\wp (z)$ is to be defined. It is clear that two derivatives of
$\rho_{n}(x)$ will be involved.  It is natural to classify the resulting
functions according to their $SO(n)$ transformation properties.
One has the symmetric traceless traceless second rank $SO(n)$ tensor
\begin{equation}
\wp ^{(n)}_{\mu \nu}\equiv(\partial_{\mu}\partial_{\nu}
-\frac{\delta_{\mu \nu}}{n}\partial_{\lambda}
\partial_{\lambda})\,\rho_{n}(x)
\end{equation}
and the $SO(n)$ scalar
\begin{equation}
\pi^{(n)}\equiv\partial_{\lambda}\partial_{\lambda}\,\rho_{n}
=-\Omega_{n}\sum..\sum\delta (x-q)
\end{equation}
already encountered in (29).  Although both (35) and (36) are
fully n-tuply periodic, the former involves true functions while the latter
is an n-fold sum over distributions.  $\wp ^{(n)}_{\mu \nu}$ is also similar
to $\wp (z)$ in that
\begin{equation}
\oint_{\partial\cal{C}}d\,\sigma_{\mu}\wp ^{(n)}_{\mu \nu}=\int_{\cal{C}}
dV_{n}\,\partial_{\mu}\,\wp ^{(n)}_{\mu \nu}=0
\end{equation}
just like $\wp (z)$ which obeys
\begin{equation}
\oint_{\partial\cal{C}}\wp (z) dz=0.
\end{equation}

\vspace{2cm}

I am grateful to A. Aliyev, O. F. Dayi, C. Delale and Y. Nutku for
patiently instructing me in the use of LATEX and to my colleagues
at the MRC for their hospitality.

\pagebreak

\end{document}